\documentclass[twocolumn]{emulateapj}
\submitted{Science with Parkes @ 50 Years Young, 31 Oct. -- 4 Nov., 2011}
\usepackage{natbib}
\usepackage{rotating}
\usepackage{mathptmx}

\newcommand{\ionhy}{H{\sc ii}}

\shorttitle{A Timeline for Star Formation}
\shortauthors{S. P. Ellingsen et al.}

\begin{document}

\title{An Evolutionary Timeline for High-mass Star Formation}

\author{S. P. Ellingsen}
\affil{School of Mathematics and Physics, University of  Tasmania, 
  Private Bag 37, Hobart, TAS 7001, Australia}
\email{Simon.Ellingsen@utas.edu.au}
\author{S.L. Breen, M.A. Voronkov, J.L. Caswell}
\affil{CSIRO Astronomy and Space Science, Australia Telescope National Facility, PO Box 76, Epping, NSW 1710, Australia}
\author{X. Chen}
\affil{Kay Laboratory for Research in Galaxies and Cosmology, Shanghai Astronomical Observatory, Chinese Academy of Sciences, Shanghai 200030, China}
\author{A. Titmarsh}
\affil{School of Mathematics and Physics, University of  Tasmania, 
  Private Bag 37, Hobart, TAS 7001, Australia}
  
\begin{abstract}
The details of the physical process through which high-mass stars form remains nearly as much of a mystery now as it was when the Parkes radio telescope commenced operation.  The energy output from high-mass stars influence, or directly drive many important processes in the evolution of galaxies and so understanding in detail when and how they form is important for a broad range of fields of astrophysics.  Interstellar masers are one of the most readily observed signposts of regions where young high-mass stars have formed.  We have recently made great progress towards using the different maser species and transitions to construct a maser-based evolutionary timeline for high-mass star formation.  Here we give an overview of this work, highlighting the particular contribution that past and on-going observations with the Parkes 64m radio telescope have made to this area.
\end{abstract}


\section{Introduction}

High-mass stars represent a very small fraction of the number of stars in the Milky Way (there are only around 3 O stars for every 100000 stars in the Galaxy), yet they produce a significant fraction of the energy output ($\sim$ 15\%).  High-mass stars play an important role in regulating the efficiency of star formation in clusters, as their ultraviolet radiation, powerful outflows and stellar winds disrupt and disperse the molecular material from which the stars are forming.  They produce heavy elements at the end of their lives when they explode as supernovae, which can both disrupt and trigger further episodes of star formation.  Despite their importance, we do not currently understand the details of the processes which enable high-mass stars to form.  One of the major limitations in advancing our understanding is that we have yet to identify a single proto-O star in the Galaxy (although there are a number of known proto-B stars).  Identifying the youngest high-mass protostars is critical to enable us to test theoretical models through comparison with observation, and here we outline recent progress towards using interstellar masers to do that.

\begin{figure*}
\begin{center}
      \includegraphics[width=120mm]{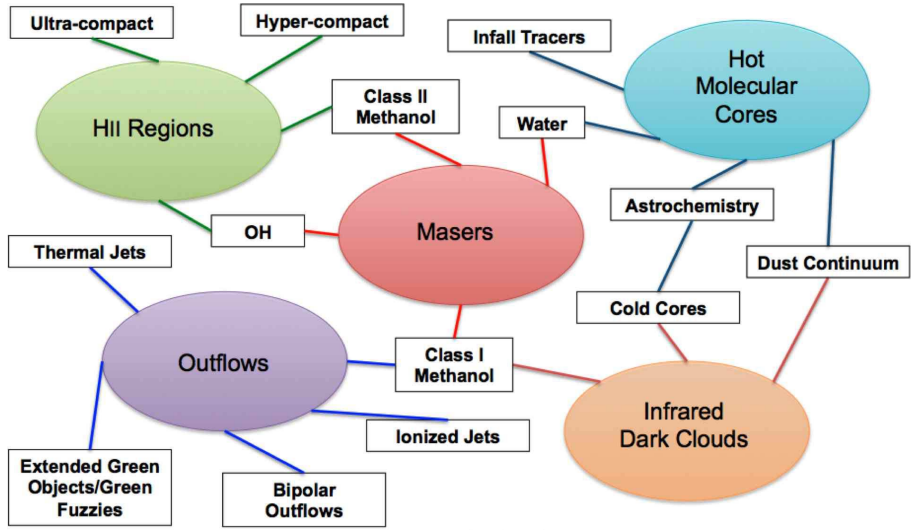}
\end{center}
\caption{A rogues gallery of astrophysical phenomena associated with high-mass star formation and the association of different types of maser transitions with them.} \label{fig:rogues}
\end{figure*}

The first interstellar masers were serendipitously detected towards young high-mass stars in searches for absorption from the OH ground-state towards the strong radio continuum emission from H{\sc ii} regions (Weaver et al. 1965).  Although maser emission has subsequently been detected from a range of other molecules and in a variety of astrophysical environments, the strongest Galactic masers are those associated with young high-mass stars.  

Regions where young, high-mass stars are forming contain warm, turbulent molecular gas, plus the presence of both a strong radiation field and shocks.  These conditions are ideal for producing population inversions in large volumes of gas.  It is thought that we observe a maser ``spot'' at locations within this gas where, by chance there is a path with a high degree of velocity coherence along our line-of-sight.  High-resolution observations show that masers occur in clusters with linear scales of around 0.01 pc (Caswell, 1997), and this will be the approximate scale of the larger masing region.  We would expect observers along lines of sight which differ from our own to observe maser emission from the same region, of similar intensity, although the lines of sight which produce velocity coherence for them will be different from those we observe and there will be variations in intensity due to the intrinsic stochasticity of the paths.

The presence of a particular maser species within a star formation complex indicates the existence of a specific range of physical conditions within a small region of the larger site (Cragg, Sobolev, \& Godfrey, 2005).  The different maser transitions and species are favoured by different, but sometimes partially overlapping physical conditions.  Common maser transitions, such as ground-state OH masers, 6.7 GHz methanol masers and 22 GHz water masers must require conditions which arise in a large number of high-mass star formation regions and persist for an extended period.  In contrast the rare maser transitions (e.g. excited-state OH masers, rare methanol and water maser transitions), must signpost conditions which are uncommon and/or relatively short-lived.  Many star formation regions show emission from multiple maser species or transitions, while others have emission from only one species.  Ellingsen et al. (2007) suggested that if it is assumed that each maser transition arises only once in the vicinity of a high-mass young stellar object, then the presence and absence of the different maser transitions can potentially be used an a clock to trace the evolution of the region.  

A qualitative maser-based evolutionary timeline for high-mass star formation incorporating the four common maser groups seen in high-mass star formation regions was first put forward by Ellingsen et al. (2007).  Over the last 5 years we have quantified and further refined this initial timeline, here we outline the role that observations with the Parkes 64m radiotelescope have played, both in the formulation of the initial timeline, and in its subsequent refinement.

\section{Masers as Evolutionary probes} \label{sec:masers}

The youngest high-mass stars are relatively distant (typically $> 1$ kpc) and heavily obscured by surrounding gas and dust at most wavelengths.  Because of this most studies of high-mass star formation focus on more readily observed associated phenomena, such as the emission from ionised gas, warm dust or from molecules at radio and millimetre wavelengths.  Each of these different tracers of high-mass star formation reveals something about the process through which high-mass stars form, however, the relationship between the various phenomena, the degree to which they are exclusively associated with a particular phase is largely unknown.  Many studies focus on a specific tracer, often largely ignoring the broader question as to where that tracer fits into the larger picture of understanding the process of high-mass star formation.

Some types of interstellar masers are very commonly observed towards high-mass star formation regions.  In particular ground-state OH masers at 1.6 GHz, methanol masers at 6.7 \& 12.2 GHz (so-called class II methanol masers), 22 GHz water masers and methanol masers at 36 and 44 GHz (class I methanol masers).  Each of these type of maser have been detected towards more than 200 sites within the Milky Way.  These different types of masers are commonly found to be associated with other phenomena associated with high-mass star formation regions (see Figure~\ref{fig:rogues}).  For example OH and class I methanol masers are often seen to be associated with ultra compact \ionhy\/ regions (Forster \& Caswell, 1989; Voronkov et al. 2010), in contrast most class II methanol and water masers usually have no centimetre continuum emission stronger than 1 mJy 	(Phillips et al. 1998; Walsh et al. 1998).  Most (perhaps all) methanol masers are associated with mid-infrared sources detectable at 24~$\mu$m, and all are associated with millimetre dust continuum emission (Ellingsen, 2006; Pandian et at. 2010; Xu et al., 2009).  Many maser sources are associated with infrared dark clouds (IRDC) and/or extended green objects (EGO) (Chen, Ellingsen, \& Shen, 2009; Chen et al. 2011; Cyganowski et al. 2009; Ellingsen, 2006), and most maser sources are likely to be at the hot-core, or earlier evolutionary phases.  This means that an evolutionary timeline based on masers can be utilised to study other aspects of high-mass star formation, which are less easy to search for in large-scale blind surveys due to sensitivity or other limitations.  A maser-based timeline has the potential to enable us to determine the relative evolutionary phase associated with each of the different tracers of high-mass star formation.

\begin{figure}
\begin{center}
\resizebox{\hsize}{!}{\includegraphics{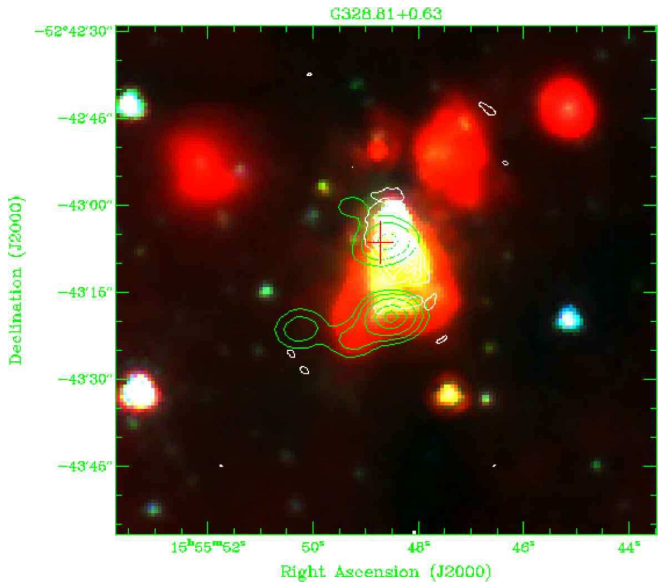}}
\end{center}
\caption{A GLIMPSE 3 colour image of the G328.81+0.63 star formation region.  The green contours are the 95 GHz class I methanol maser emission and the red cross marks the location of the class II methanol, OH and water masers.} \label{fig:g328}
\end{figure}

\begin{figure}
\begin{center}
\resizebox{\hsize}{!}{\includegraphics{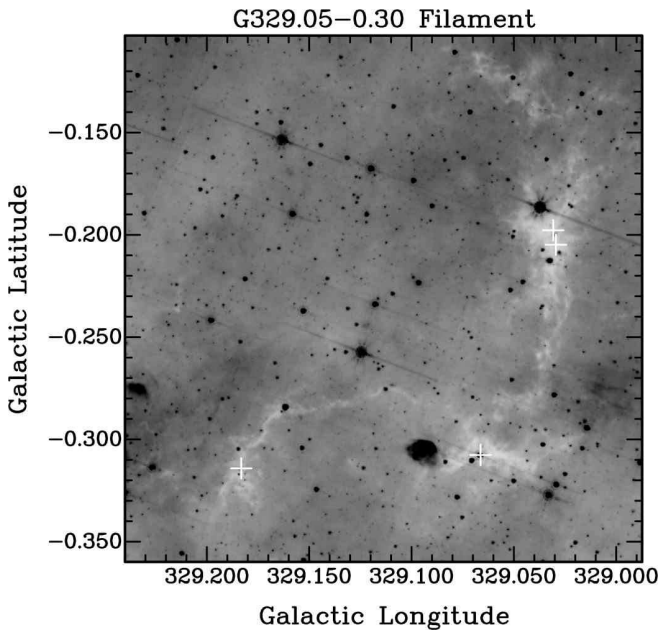}}
\end{center}
\caption{An 8$\mu$m GLIMPSE image of the G329.05-0.30 filamentary infrared dark cloud.  The image is shown in reverse video, so the dark cloud appears white.  The white crosses mark the locations of the 6.7 GHz methanol maser sources G329.021-0.198, G329.029-0.025, G329.066-0.308, G329.183-0.314 (Ellingsen, 2006)} \label{fig:g329}
\end{figure}

{\em IRAS}16547-4247 (also known as G343.12-0.06) provides an example of the wide range of phenomena which are observed towards some high-mass star formation regions.  This source is well known for having a radio jet (Brooks et al. 2007), a number of rare class I methanol masers (Voronkov et al., 2006), water and OH masers, but no class II methanol masers (Breen et al. 2010a) and an molecular hydrogen outflow (Brooks et al., 2003).  Figure~\ref{fig:g328} shows the G328.81+0.63 star formation region, which shows emission from all four types of common masers, projected onto, or around a bright mid-infrared source associated with an ultracompact \ionhy\/ region.  Figure~\ref{fig:g329} shows the G329.03-0.20 star formation region, which shows multiple maser sites, all associated with one very elongated IRDC (Ellingsen, 2006).  Each of the four maser sites has OH, water, class I and class II methanol maser emission from the same general region (Breen et al., 2010a; Ellingsen, 2005).  These specific examples give a small taste of the diversity of the regions traced by the various common maser transitions.  While much can be learned from detailed studies of individual regions such as these, to obtain a more reliable and accurate picture of the relationship between the various high-mass star formation tracers we need use large samples, preferably drawn from statistically complete observations.

\begin{figure}
\begin{center}
\resizebox{\hsize}{!}{\includegraphics{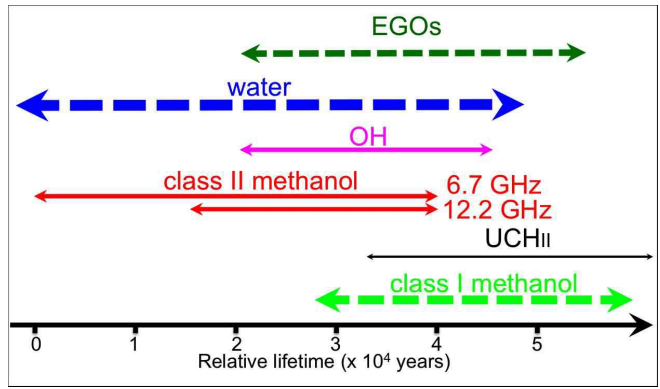}}
\end{center}
\caption{A maser-based evolutionary timeline for high-mass star formation, based on Breen et al (2010b).} \label{fig:timeline}
\end{figure}

\section{The Maser-based timeline: A Legacy of Parkes surveys}

Figure~\ref{fig:timeline} shows the maser-based evolutionary timeline we have developed, as it stands at present.  This has been derived by comparing the rates at which the different types of masers and other star-formation tracers are associated with each other, and also by looking at how the properties of the masers themselves vary within association groups.  For example Breen et al., (2011) have shown that 12.2 GHz methanol masers are generally associated with more luminous 6.7 GHz masers and that OH masers are associated with only the most luminous 6.7 and 12.2 GHz methanol masers.  They also show that the velocity range exhibited by the different maser transitions increases with increasing luminosity, suggesting that the volume of gas where conditions are suitable for masers to arise increases as the source evolves.  In addition to the common class II methanol masers, there are a number of rare transitions at higher frequencies.  One of these is the 37.7 GHz transition, the first southern search for which was recently undertaken with the Mopra telescope.  Comparison of the properties of the sources which do exhibit 37.7 GHz methanol masers, with those that do not shows that they are associated with only the most luminous 6.7 GHz methanol masers.  These masers are thought to be the most evolved of the class II methanol masers, suggesting that the 37.7 GHz masers mark the last 1000-4000 years of this maser phase in high-mass star formation regions (Ellingsen et al., 2011).

Our ability to develop a maser-based evolutionary timeline for high-mass star formation has been largely based on observations made with the Parkes telescope.  In order to be able to undertake useful statistical comparison of the different maser species requires sensitive, large-scale, unbiased searches.  The two major maser surveys which fulfil these criteria are the Parkes ground-state OH maser survey undertaken by Jim Caswell \& collaborators in the 1980s (Caswell \& Haynes, 1983; 1987; Caswell, Haynes, \& Goss, 1980), and the more recent 6.7 GHz Methanol Multibeam (MMB) survey (Caswell et al., 2010; 2011; Green et al., 2010; 2012).  These surveys have been supplemented with data from public legacy surveys such as GLIMPSE, and targeted follow-up searches of for water masers (Breen et al., 2010a) and other less common maser transitions (Breen et al., 2011; 2012).  

\section{Conclusions}

Future, similar searches for water masers (e.g. HOPS (Walsh et al., 2011)) and class I methanol masers, along with complementary information from radio through sub-millimetre continuum observations and thermal line observations (e.g. MALT90) will enable us to further refine and quantify the timeline.  The Parkes telescope may play a role in one or more of these possible future searches, further building upon its already impressive legacy.  For example a 22 GHz multibeam receiver has been discussed as a possible future instrument for Parkes.  Such an instrument would invaluable in enabling a search for water masers, thermal ammonia and possibly a range of other rarer thermal and maser lines approximately an order of magnitude more sensitive than HOPS, and comparable to the sensitivity of the MMB survey.

\section*{Acknowledgements}

The Australia Telescope Compact Array is part of the Australia Telescope which is funded
by the Commonwealth of Australia for operation as a National Facility
managed by CSIRO.  This research has made use of NASA's Astrophysics Data 
System Abstract Service.


\begin{references}

\reference{}Breen, S. L., Caswell, J. L., Ellingsen, S. P., \& Phillips, C. J. 2010a, MNRAS, 406, 1487 
\reference{}Breen, S. L., Ellingsen, S. P., Caswell, J. L., \& Lewis, B. E. 2010b, MNRAS, 401, 2219 
\reference{}Breen, S. L., Ellingsen, S. P., Caswell, J. L., Green, J. A., Fuller, G. A., Voronkov, M. A., Quinn, L. J., et al., 2011, ApJ, 733, 80 
\reference{}Breen, S. L., Ellingsen, S. P., Caswell, J. L., et al. 2012, MNRAS, 421, 1703 
\reference{}Brooks, K. J., Garay, G., Mardones, D., \& Bronfman, L. 2003, ApJ, 594, L131
\reference{}Brooks, K. J., Garay, G., Voronkov, M., \& Rodríguez, L. F. 2007, ApJ, 669, 459 
\reference{}Caswell, J. L. 1997, MNRAS, 289, 203 
\reference{}Caswell, J. L., \& Haynes, R. F. 1983, Australian Journal of Physics, 36, 361 
\reference{}Caswell, J. L., \& Haynes, R. F. 1987, Australian Journal of Physics, 40, 215, 238 
\reference{}Caswell, J. L., Fuller, G. A., Green, J. A., et al. 2010. MNRAS, 404, 1029 
\reference{}Caswell, J. L., Fuller, G. A., Green, J. A., Avison, A., Breen, S. L., Ellingsen, S. P., Gray, M. D., et al. 2011,MNRAS, 417, 1964 
\reference{}Caswell, J. L., Haynes, R. F., \& Goss, W. M. 1980, Australian Journal of Physics, 33, 639 
\reference{}Chen, X., Ellingsen, S. P., \& Shen, Z. Q., 2009, MNRAS, 396, 1603 
\reference{}Chen, X., Ellingsen, S. P., Shen, Z.-Q., Titmarsh, A., \& Gan, C.-G., 2011,ApJ Supplement Series, 196, 9 
\reference{}Cragg, D. M., Sobolev, A. M., \& Godfrey, P. D. 2005, MNRAS, 360, 533 
\reference{}Cyganowski, C. J., Brogan, C. L., Hunter, T. R., \& Churchwell, E. 2009,ApJ, 702, 1615 
\reference{}Ellingsen, S. P. 2005, MNRAS, 359, 1498 
\reference{}Ellingsen, S. P. 2006, ApJ, 638, 241 
\reference{}Ellingsen, S. P., Breen, S. L., Sobolev, A. M., Voronkov, M. A., Caswell, J. L., \& Lo, N. 2011, ApJ, 742, 109 
\reference{}Ellingsen, S. P., Voronkov, M. A., Cragg, D. M., Sobolev, A. M., Breen, S. L., \& Godfrey, P. D. 2007, Astrophysical Masers and their Environments, Proceedings of IAU Symp 242, 213, Eds J. Chapman, W. Baan 
\reference{}Forster, J., \& Caswell, J. 1989. A\&A, 213, 339 
\reference{}Green, J. A., Caswell, J. L., Fuller, G. A., et al. 2010, MNRAS, 409, 913 
\reference{}Green, J. A., Caswell, J. L., Fuller, G. A., et al. 2012, MNRAS, 420, 3108 
\reference{}Pandian, J. D., Momjian, E., Xu, Y., Menten, K. M., \& Goldsmith, P. F., 2010, A\&A, 522, A8 
\reference{}Phillips, C. J., Norris, R. P., Ellingsen, S. P., \& McCulloch, P. M. 1998, MNRAS, 300, 1131 
\reference{}Voronkov, M. A., Brooks, K. J., Sobolev, A. M., Ellingsen, S. P., Ostrovskii, A. B., \& Caswell, J. L., 2006, MNRAS, 373, 411 
\reference{}Voronkov, M. A., Caswell, J. L., Ellingsen, S. P., \& Sobolev, A. M., 2010, MNRAS, 405, 2471  
\reference{}Walsh, A. J., Breen, S. L., Britton, T. et al. 2011, MNRAS, 416, 1764 
\reference{}Walsh, A., Burton, M., Hyland, A., \& Robinson, G. 1998, MNRAS, 301, 640 
\reference{}Xu, Y., Voronkov, M. A., Pandian, J. D., et al. 2009, A\&A, 507, 1117 
\reference{}Weaver, H., Williams, D.R.W., Dieter, N.H., Lumn, W.T., 1965, Nat, 208, 29 

\end{references}
\end{document}